\documentclass[10pt, conference, compsocconf]{IEEEtran}
\usepackage{multicol}
\usepackage{a4wide}
\usepackage{multirow}
\usepackage{caption}
\usepackage{subcaption}
\usepackage{amsfonts}
\usepackage{mathtools}
\usepackage{bm}
\usepackage{tabu}

\usepackage[table]{xcolor}

\newtheorem{mydef}{Definition}[section]
\usepackage[font=footnotesize, labelfont={bf}]{caption}

\begin{document}

\title{Network comparison using directed graphlets}

\author{David Apar\'{i}cio \and Pedro Ribeiro \and Fernando Silva}

\author{
\IEEEauthorblockN{David Apar\'{i}cio, Pedro Ribeiro, Fernando Silva}
\IEEEauthorblockA{CRACS \& INESC-TEC LA, \\
  Faculdade de Ci\^encias, Universidade do Porto \\
  R. Campo Alegre, 1021/1055, 4169-007 Porto, Portugal \\
Email: \{daparicio, pribeiro, fds\}@dcc.fc.up.pt}
}

\maketitle

\begin{abstract}
With recent advances in high-throughput cell biology the amount of
cellular biological data has grown drastically. Such data is often
modeled as graphs (also called networks) and studying them
 can lead to new
insights into molecule-level organization. A possible way to
understand their structure is by analysing the smaller
components that constitute them, namely network motifs and graphlets. Graphlets are
particularly well suited to compare networks and to assess their level
of similarity but are almost always used as small undirected graphs of
up to five nodes, thus limiting their applicability in directed
networks. However, a large set of interesting biological networks such
as metabolic, cell signaling or transcriptional regulatory networks
are intrinsically directional, and using metrics that ignore edge
direction may gravely hinder information extraction.  

The applicability of graphlets is extended to directed networks by considering the edge direction of the graphlets. We tested our approach on a set of directed biological networks and verified that they were correctly grouped by type using directed graphlets. However, enumerating all graphlets in a large network is a computationally demanding task. Our implementation addresses this concern by using a state-of-the-art data structure, the g-trie, which is able to greatly reduce the necessary computation.  We compared our tool, \texttt{gtrieScanner}, to other state-of-the art methods and verified that it is the fastest general tool for graphlet counting.
\end{abstract}

\section{Introduction}

The advent of high-throughput cell biology technologies such as DNA
microarrays \cite{schena1995quantitative} has increased the amount of
data pertaining to molecular interactions exponentially. While this
recent flood of information has greatly contributed to a more accurate
understanding of molecule-level organization, it has also created the
need to find ways to filter and model this data so that it is rendered
intelligible to the practitioner. In the field of computational
biology different types of cellular networks are modeled as graphs
with nodes representing specific biological components such as
proteins or genes, and the physical, chemical or functional
interactions between them modeled as edges.

Inspecting the graph's topological features can yield valuable
information about the network. If a network has topological features
that are not expected to occur in neither purely random nor purely
regular graphs it is considered to be a complex network, and most
real-world networks are found to be complex. Singular characteristics
commonly associated with complex networks include having a small-world
structure \cite{wagner2001small} or a degree distribution that
follows a power-law (scale-free networks)
\cite{jeong2000large}. 
Brain networks, for instance, have been
identified as small-world networks \cite{sporns2004organization},
meaning that each node (representing a neuron) is only a few
connections away from every other node. Furthermore, the distance of
the average path length in the brain has been negatively correlated
with a person's IQ \cite{van2009efficiency}, which seems to indicate
the importance of a small-world organization to networks'
efficiency. 
Other statistics such as the number of connected
components and the clustering coefficient are also frequently used to
characterize a network. 

Another approach to uncover the underlying structure of complex
networks is to decompose them into their smaller components or
\textit{subgraphs}. Network motifs in particular are small
overrepresented subgraphs described by \cite{Milo2002} as the building
blocks of large complex networks. Studies such as the one by
\cite{prill2005dynamic} support this view for biological networks and
network motifs have been used to analyze a wide range of cellular
biological networks such as protein-protein interaction networks
\cite{albert2004conserved}, transcriptional regulatory networks
\cite{shen2002network}, metabolism networks
\cite{shellman2013network} or cell signaling networks
\cite{li2013cancer}.

It is often useful to compare networks against each other,
particularly because if a given network's properties are known it
allows for knowledge transfer based on the similarity or difference
between two networks. One way to perform this comparison is to
evaluate the similarity between the subgraphs that each network
contains. Network motif fingerprints \cite{alon2004} and graphlet-based metrics \cite{Przulj2007} are possible choices for this task. Both
approaches compute the frequency of a set of small non-isomorphic subgraphs but, in addition to that, graphlets also evaluate the contribution of each individual node from the network, producing a graphlet degree distribution that can be seen as an extension of the node degree concept. Furthermore, enumerating graphlets is computationally less expensive than calculating network motifs since the subgraphs are only enumerated in the original network, as opposed to also having to compute their occurrences in a large set of random networks in order to assess motif
significance \cite{ribeiro2009}. Graphlet-based metrics have
been used to analyze networks from various biological areas such as
protein-protein interaction \cite{kuchaiev2009learning}, disease
genes \cite{sun2014predicting}, age-related genes
\cite{yoo2014improving} or brain networks
\cite{kuchaiev2009structure} due to the great amount of topological information that they
provide \cite{yaverouglu2015proper}.


Graphlet usage is often restricted to analyzing only the set of 30
undirected graphs of up to five nodes originally presented by
\cite{Przulj2007} due to computational limitations. However, a
substantial gain in topological information might be attained by
examining larger graphlets. For instance, two networks may
be almost indistinguishable by comparing their respective size
$k$ graphlets but certain dissimilarities become evident when contrasting 
their size $k+1$ graphlets. Additionally, in directed networks, edge 
direction should be taken into account since it can potentially reveal
information about the network's structure that undirected
graphlets are not able to capture. In this work we propose a
novel extension of the graphlet methodology to directed networks and
show how this may retrieve relevant information from biological
networks. Despite the fact that graphlets are a general model, to the
best of the authors' knowledge the only extensions to the original
concept are relative to ordered \cite{Malod-Dognin2014} and 
dynamic graphs \cite{hulovatyy2014exploring}.

There are many cellular biological networks that are intrinsically
directed such as metabolic, cell signaling and gene transcriptional
regulation networks. Methods and metrics that ignore the edge
direction of these networks might be losing important
information. \cite{Garlaschelli2004} proposed a new measure ($\rho$) to calculate the \emph{link
reciprocity} of a network that can be used to assess if its edge direction
is important or not. Their measure is an absolute quantity ranging from $-1$ (\emph{no reciprocity})
to $1$ (\emph{completely reciprocal}). Networks with a $\rho$-value of $\approx-1$ 
are \textit{purely directional networks} meaning that
edge direction is an intrinsic aspect of these networks and 
removing it makes them meaningless. On the other hand, networks with $\rho
\approx 1$ can be safely transformed into topologically equivalent undirected networks 
without losing much information since their edges are always reciprocally connected. 
Garlaschelli and Loffredo calculated that celular and food web networks rank closer
to the middle of the scale ($\rho \approx 0$) meaning that edge direction  
in these  networks is significant. Additionally, some specific small directed graphs, such
as feed-forward loops, have been shown to play a fundamental role in
the organization of distinct networks \cite{mangan2003structure}.

Network motifs have been extensively used to study directed biological
networks such as neural, transcriptional and signal networks \cite{wang2014identification2}. 
Graphlets on the other hand are mostly restricted to
undirected networks since they consist of a set of undirected
graphs. \cite{park2010simultaneous} examined numerous directed
biological networks using both directed motifs and undirected
graphlets. Such a study could possibly benefit if a tool for enumerating
directed graphlets was available.

In this article we present an efficient general-purpose tool to enumerate and
compare both directed and undirected graphlet degree distributions for
again both directed and undirected networks. Previous approaches
either restricted the application of graphlets to undirected networks
or had to ignore edge direction in directed networks, in practice
reducing them to undirected networks. To achieve this objective we
extend both i) the original concept of graphlets to \emph{directed
  graphlets} and ii) upgrade a tree data-structure specialized in
efficiently storing graphs, the g-trie, to a \emph{graphlet-trie}. Our
tool, \texttt{gtrieScanner}, can thus be used
to enumerate directed and undirected graphlets as well as network
motifs.

\section{Materials and Methods}

\subsection{Graph and graphlet terminology}

A network or \textit{graph} $G$ is comprised of a set $V(G)$ of \textit{vertices} or \textit{nodes} and a set $E(G)$ of \textit{edges} or \textit{connections}. Nodes represent entities and edges correspond to the relationships between them. Edges are represented as pairs of vertices of the form $(a, b)$, where $a, b \in V(G)$. In \textit{directed} graphs, edges $(i, j)$ are \textit{ordered pairs} (translated to "$i$ \textit{goes to} $j$") whereas in \textit{undirected} graphs there is no order since the nodes are always reciprocally connected. 

A $subgraph$ $G_k$ of $G$ is a graph of size $k$ where $V(G_k) \subseteq V(G)$ and $E(G_k) \subseteq E(G)$. A subgraph is $induced$ if $\forall u, v \in V(G_k): (u,v) \in E(G_k)$ iff $(u, v) \in E(G)$. A $match$ or $occurrence$ of $G_k$ happens when $G$ has a set of nodes that induce $G_k$. Two matches are considered distinct if they have at least one different vertex.  The $frequency$ of $G_k$ in $G$ is the number of occurrences of $G_k$ in $G$. 

Two graphs are said to be \emph{isomorphic} if it is possible to obtain one from the other by changing the node labels without affecting their topology. All occurrences of a set $\mathcal{G}$ of non-isomorphic subgraphs must be enumerated in the original network before graphlet or network motif metrics can be calculated. We call this task the general subgraph census problem and state it in Definition~\ref{def:genproblem}.

\begin{mydef}[\textbf{Subgraph Census Problem}]
\label{def:genproblem}
Given a set $\mathcal{G}$ of non-isomorphic subgraphs and a graph $G$, determine the frequency of all induced occurrences of the subgraphs $G_s \in \mathcal{G}$ in $G$. Two occurrences are considered different if they have at least one node or edge that they do not share. Other nodes and edges can overlap.
\end{mydef}

Graphlets were originally defined as the set of 30 non-isomorphic undirected graphs with at most 5 nodes \cite{Przulj2007}. They are structurally equivalent to \textit{network motifs} \cite{Milo2002} but also include information about the position or \textit{orbit} that nodes occupy in the graphlet. The set of all orbits of $\mathcal{G}$ is refereed to as $\mathcal{O}$. Another difference lies in the fact that network motifs require a null model to verify if the subgraph is overrepresented. Usually the null model is an artificial random network that maintains the original network's node degree sequence. On the other hand, graphlets do not require a null model and use the information of all subgraphs to perform a full-scale network comparison. 

To compute the graphlet degree distribution one has to count $\forall u \in V(G)$ how many times $u$ appears in some orbit $j \in \mathcal{O}$ and repeat this process for the total $m$ orbits, resulting in a graphlet degree vector $GDV(u)$ with $m$ positions. A matrix $Fr_G$ of $n \times m$ is obtained by joining the $GDV$s of all $n$ nodes where each row of $Fr_G$ is $GDV(v), v \in V(G)$ and each position $fr_{u,j}$ is the number of times that node $u$ appears in orbit $j$. This task is more formally defined in Definition~\ref{def:genproblem2}.

\begin{center}
$Fr_G =
 \begin{bmatrix}
  fr_{0,0} & fr_{0,1} & \cdots & fr_{0,m} \\
  fr_{1,0} & fr_{1,1} & \cdots & fr_{1,m} \\
  \vdots  & \vdots  & \ddots & \vdots  \\
  fr_{n,0} & fr_{n,1} & \cdots & fr_{n,m}
 \end{bmatrix}$
\end{center}

\begin{figure}
  \centering
  \includegraphics[width=0.9\linewidth]{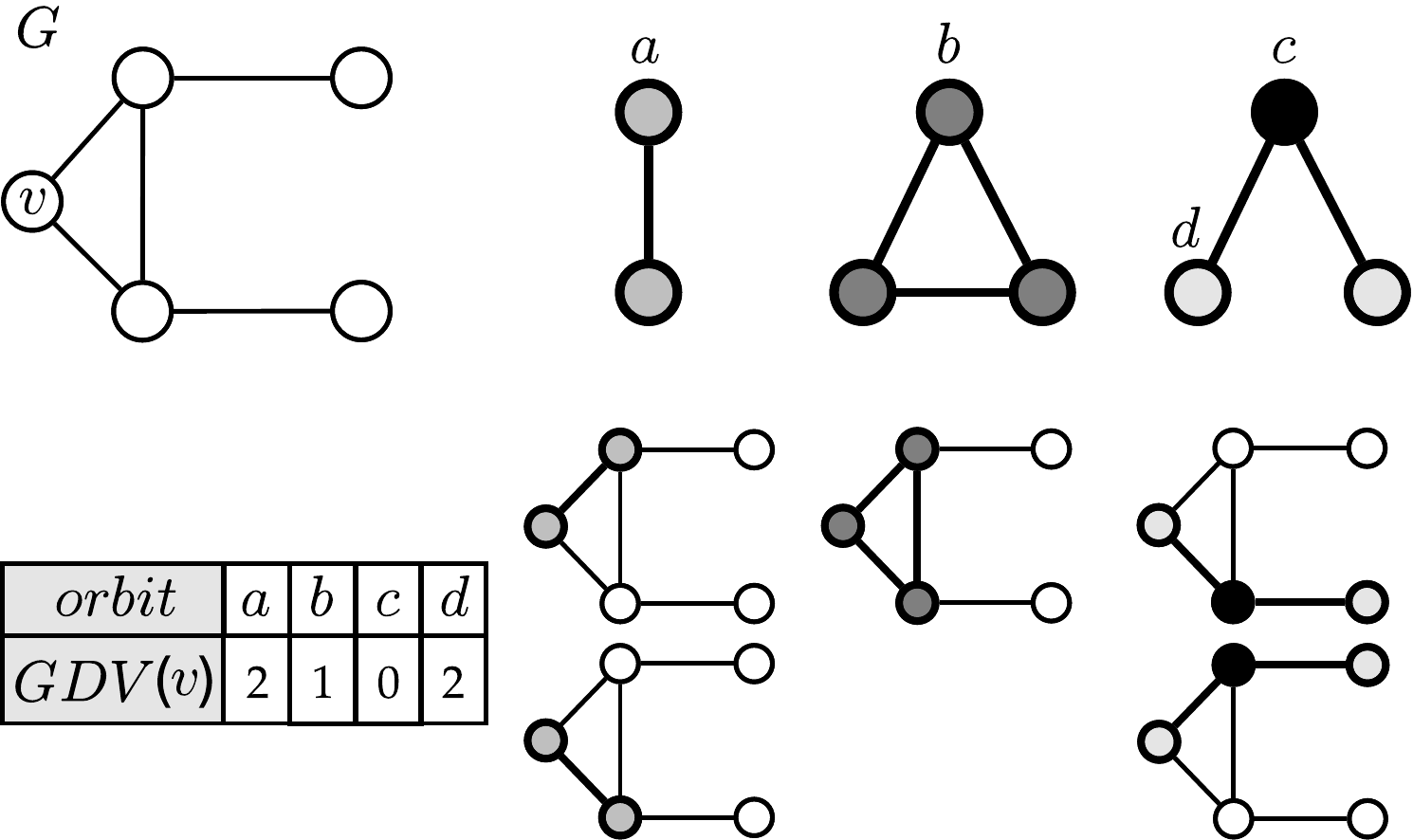}
  \caption{$GDV(v)$ obtained by enumerating $v$'s undirected graphlet orbits of sizes 2 and 3.}
  \label{fig:occ_orbits}
\end{figure}

Both directed and undirected graphlets are considered in this work; $d\mathcal{G}_k$ and $d\mathcal{O}_k$ denote the set of all directed graphlets and directed orbits of size $s\in \left\{2, ..., k\right\}$, respectively, while $u\mathcal{G}_k$ and $u\mathcal{O}_k$ are used for the undirected counterparts. An illustration of how the $GDV$ of a vertex $v$ is computed for $u\mathcal{O}_3$ is presented in Figure~\ref{fig:occ_orbits}.

\vspace{0.4cm}

\begin{mydef}[\textbf{Orbit Frequency Computation}]
\label{def:genproblem2}
Given a set $\mathcal{G}_s$ of non-isomorphic subgraphs of size $s \in \left\{2, ..., k\right\}$ and a graph $G$, determine the number of times $fr_{i,j}$ that each node $i \in V(G)$ appears in all the orbits $j \in \mathcal{O}_s$. Two occurrences are considered different if they have at least one node or edge that they do not share. Other nodes and edges can overlap.
\end{mydef}
 
$Fr_G$ is transformed into a graphlet degree distribution $GDD_G$ where $d_G^{\,j}(k)$ denotes how many nodes appear $k$ times in orbit $j$.
 
\begin{center}
 $GDD_G =
  \begin{bmatrix}
   d_G^{\,0}(1) & d_G^{\,0}(2) & \cdots & d_G^{\,0}(\scalebox{0.7}{\!+}\infty) \\
   d_G^{\,1}(1) & d_G^{\,1}(2) & \cdots & d_G^{\,1}(\scalebox{0.7}{\!+}\infty) \\
   \vdots  & \vdots  & \ddots & \vdots  \\
   d_G^{\,m}(1) & d_G^{\,m}(2) & \cdots & d_G^{\,m}(\scalebox{0.7}{\!+}\infty)
  \end{bmatrix}$
\end{center}

\begin{table*}[t] 
\centering
\footnotesize
\begin{tabular}{|c||c|c||c|c|c|}
\cline{2-5}

\multicolumn{1}{c|}{}& \multicolumn{2}{c||}{$u\mathcal{G}_k$} & \multicolumn{2}{c|}{$d\mathcal{G}_k$}     \\  \hline
$k$       & $|\mathcal{G}_k|$ & $|\mathcal{O}_k|$     & $|\mathcal{G}_k|$ & $|\mathcal{O}_k|$ \\  \hline  \hline
2         & 1         & 1              & 2         & 3  (1.5  $\times$ $|\mathcal{G}_k|$)    \\  \hline
3        & 3         & 4             & 15        & 33  (2.3  $\times$ $|\mathcal{G}_k|$)   \\  \hline
4      & 9         & 15         & 214       & 730  (3.5  $\times$ $|\mathcal{G}_k|$)  \\  \hline
5    & 30        & 73        & 9,578      & 45,637  (4.8  $\times$ $|\mathcal{G}_k|$)\\  \hline
6  & 142       & 480      &    1,540,421       &   9,121,657  (5.9  $\times$ $|\mathcal{G}_k|$)   \\  \hline
7  & 965       & 4,786              &      882,011,563     &    $\approx$ 7 $\times$ $|\mathcal{G}_k|$    \\  \hline
8  & 12,082     & 77,275             &        1,793,355,966,869   &  $\approx$ 8 $\times$ $|\mathcal{G}_k|$    \\  \hline
9   & 273,162    & 2,188,288           &       13,027,955,038,433,121    & $\approx$ 9 $\times$ $|\mathcal{G}_k|$    \\   \hline
\end{tabular}
\caption{Number of undirected and directed graphlets, as well as their respective orbits, depending on the size of the graphlets. For each case, all graphlets of sizes ${2..k}$ are counted. It is impractical to enumerate all possible orbits for $d\mathcal{G}_k$ when $k$ is larger than 6 due to the size of $|\mathcal{G}|$. \label{table:orbits}}
\end{table*}
  
Two networks $G$ and $H$ can then be compared by computing the differences between their respective $GDD$ matrices after their distributions are normalized - represented below as $n^j_G(k)$. In our experiments the arithmetic mean GDD-agreement ($GDA$) introduced by \cite{Przulj2007} is used, defined as follows.

\begin{equation}
GDA(G,H)^j = 1 - \frac{1}{\sqrt{2}} \left(\sum\limits_{k=1}^{\scalebox{0.6}{\!+}\infty}[n^j_G(k) - n^j_H(k)]^2\right) ^\frac{1}{2}
\end{equation}

\begin{equation}
GDA(G,H) = \frac{1}{m}\sum\limits_{j=0}^{m}GDA(G,H)^j
\end{equation}

\begin{figure}[!ht]
   \centering
   \includegraphics[width=0.35\textwidth]{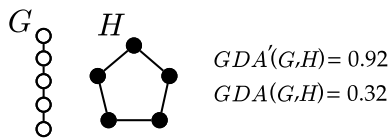}
   \caption{Comparison of different $GDA$ metrics. The original metric $GDA'$ was found to produce unreasonably high values for small networks (in the example) or when many graphlets are enumerated. This makes the metric inappropriate to compare directed graphlets since the number of orbits is very high. In our modified $GDA$ metric only orbits appearing in either $G$, $H$ or both are considered, discarding non-present orbits.}
   \label{fig:gdd_comp}
\end{figure}

We modified the metric to only consider orbits that appear in at least one of the networks; the original metric is henceforth refereed to as $GDA'$ and our own as $GDA$. Modifying the metric was necessary since non-appearing orbits would contribute to unreasonably high $GDA'$s when enumerating a large number of orbits or when small networks were used. This happens because the $GDA'$ of two networks is increased even if the orbit frequency is zero in both networks. This is not very problematic when a small number of orbits is enumerated, such as the original 73 undirected ones; however, as can be seen from Table~\ref{table:orbits}, bigger undirected graphlets and directed graphlets may require thousands or millions of orbits to be enumerated. For these cases it is likely that many of the possible orbits do not appear in either network, which may result in higher $GDA'$s than expected. Tests performed on small food webs produced an average $GDA'$ of $\approx 0.5$ when enumerating $d\mathcal{G}_4$ that increased to $\approx0.85$ for $d\mathcal{G}_5$. This does not translate to those foodwebs being much more similar when looking at their larger graphlets but rather that there are many orbits that do not appear in either network. Figure~\ref{fig:gdd_comp} illustrates this concept. Since both directed and undirected $GDA$s can be calculated, $uGDA_k$ and $dGDA_k$ represent the $GDA$s when comparing undirected and directed graphlets, respectively, of size $k$.

\begin{figure*}[!ht]
   \centering
   \includegraphics[width=0.6\textwidth]{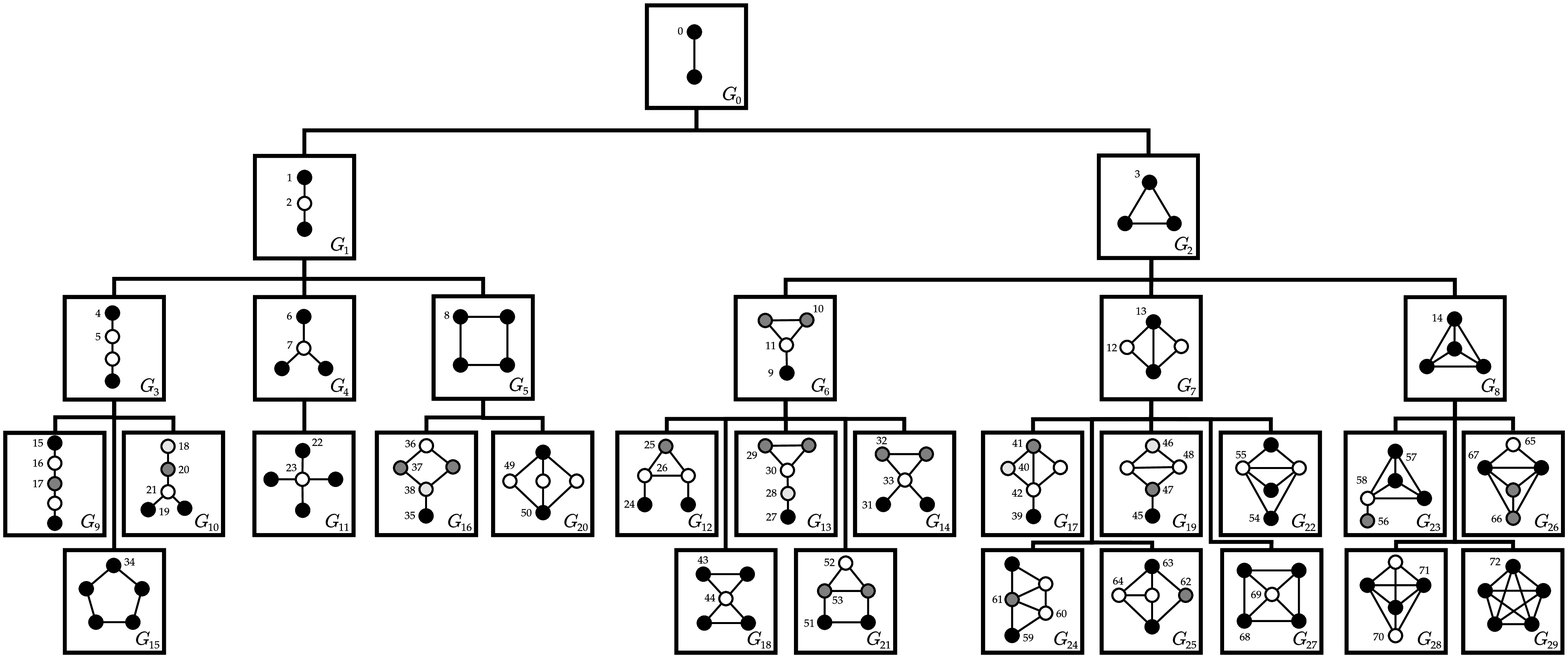}
   \caption{A graphlet-trie containing all 2, 3, 4 and 5-node undirected graphlets $G_0, ..., G_{29}$ as they were presented in \cite{Przulj2007}. In a g-trie the common topologies between graph(let)s of different sizes become evident. }
   \label{fig:graphlettrie}
\end{figure*}

\begin{figure*}[!ht]
   \centering
   \includegraphics[width=1.0\textwidth]{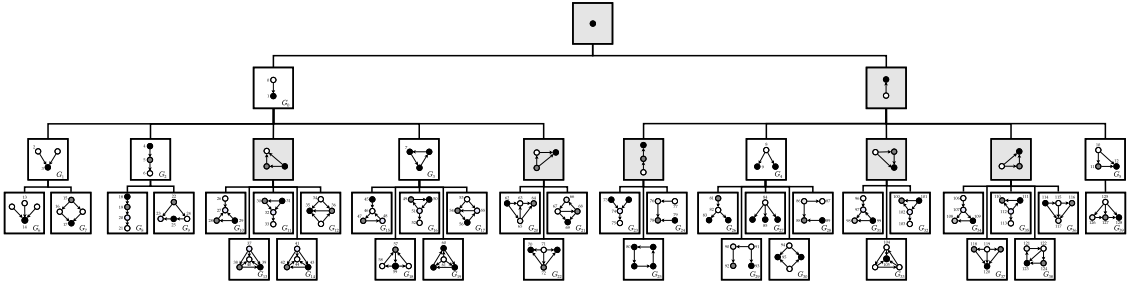}
   \caption{A graphlet-trie containing the 39 non-bidirectional directed graphlets of sizes 2, 3 and 4. The orbit numbers are generated automatically. While it is guaranteed that the g-trie only has non-isomorphic graphs on the leafs (bottom-level nodes), isomorphic graphs may appear in some of the top nodes (represented in grey). In these cases only one of the graphs is considered for orbit counting while the others are only used to efficiently traverse the search space, using symmetry breaking conditions to guarantee that each occurrence is only counted once. }
   \label{fig:graphlettrie_dir}
\end{figure*}

\subsection{Networks}

There are numerous kinds of intra-cellular networks, such as metabolic, transcriptional regulatory and cell signaling networks, where edge directyion is intrinsically related to its function.

Metabolic networks represent the set of biochemical reactions occurring within a cell that allow the organism to grow, reproduce, respond to the environment, and other biological functions essential for the organism's survival. These reactions are catalyzed by enzymes that act upon substrates. Therefore, in metabolic networks a node can be an enzyme or a substrate and the connections are directed edges going from enzymes to substrates. 

Transcriptional regulatory networks model the process by which the information in the genes is transcribed into proteins or RNA, also called gene expression. In these networks nodes are either transcription factors or proteins that are connected by directed edges representing how the transcription factors influence the gene by stimulating or repressing its expression.

A cellular signaling network is comprised of a sequence of biochemical reactions between cells of the same organism. A great number of tasks such as the development, repair and immunity of cells depend on the proper functioning of cell signaling networks. Nodes in these networks are proteins and edges exist between activator and receptor proteins that communicate through signals from the first to the latter. 

\begin{figure}[!ht]
   \centering
   \includegraphics[width=0.25\textwidth]{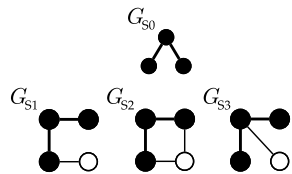}
   \caption{Common topology of three small graphs. G-Tries use common structures between the graphs of $\mathcal{G}_s$ to heavily constrain the search space.}
   \label{fig:commontop}
\end{figure}

\subsection{The G-Trie Data-structure}

A g-trie \cite{Ribeiro2014} is a tree-like data-structure created initially to calculate network motifs but it can be efficiently used to solve the general subgraph census problem. \cite{ribeiro2010g} presented a g-trie algorithm that was one or two orders of magnitude faster than previous approaches and g-tries are still state-of-the-art for network motif discovery efficiency-wise. 

The efficiency of the g-trie data-structure is mostly due to two main algorithmic ideas. First, the search space is heavily constrained by identifying common subtopologies between the set of subgraphs $\mathcal{G}_s$ before enumerating them. Figure~\ref{fig:commontop} illustrates this base concept by showing three small graphs that share a common subtopology. In practice this means that instead of searching each subgraph, $G_{S1}$, $G_{S2}$ and $G_{S3}$, individually, a g-trie starts by looking for occurrences of a smaller common subgraph $G_{S0}$ and then perform the necessary expansions for each larger subgraph. Secondly, symmetry breaking conditions are automatically generated to eliminate automorphisms, thus avoiding redundancies and guaranteeing that each occurrence is found only once. For instance, without these conditions an $n$-clique ($n$ nodes fully connected to each other) would be found $n!$ times (once for each possible permutation of the $n$ nodes) since all these permutations would be a match to a g-trie path from the root to the node representing that $n$-clique.


\begin{table*}[!ht]
\small
\centering
\addtolength{\tabcolsep}{-1.1mm}
\begin{tabu}{|c||p{1.2cm}|p{1.2cm}|p{1.2cm}|p{1.2cm}|p{1.2cm}|p{1.2cm}|p{1.2cm}|p{1.2cm}|p{1.2cm}|c|}
\hline
$G$  & {\tt SIG\_NCI}  & {\tt SIG\_NH} & {\tt SIG\_SH} & {\tt SIG\_SM} & {\tt MET\_CE} &  {\tt MET\_BS} &  {\tt MET\_DR} &  {\tt MET\_TY} &   {\tt TR\_EC} &  {\tt TR\_YST}  \\ \hline
$|V(G)|$ & 15,533 & 1,634 & 529 & 477 & 2,217 & 453  & 2,280 & 2,361 & 99 & 688 \\ \hline
$|E(G)|$ & 23,682 & 4,665 & 1,223 & 1,056 & 5,427 & 2,025 & 5,588 & 5,822  & 212 & 1,078 \\ \hline
\rowfont{\scriptsize} {\small Source} & \cite{schaefer2009pid} & \cite{cui2007map} & \multicolumn{2}{c|}{\cite{ma2009snavi}} & \cite{duch2005community} & \multicolumn{3}{c|}{\cite{jeong2000large}} &  \cite{mangan2003structure} & \cite{Milo2002}\\ \hline
\end{tabu}
\caption{Set of ten directed biological networks of different types used for experimental evaluation. \label{tab:networks} }

\addtolength{\tabcolsep}{+0.5mm}
\end{table*}

\begin{figure*}[!ht]
   \centering
   \vspace{-2.0cm}
   \includegraphics[width=1.0\textwidth]{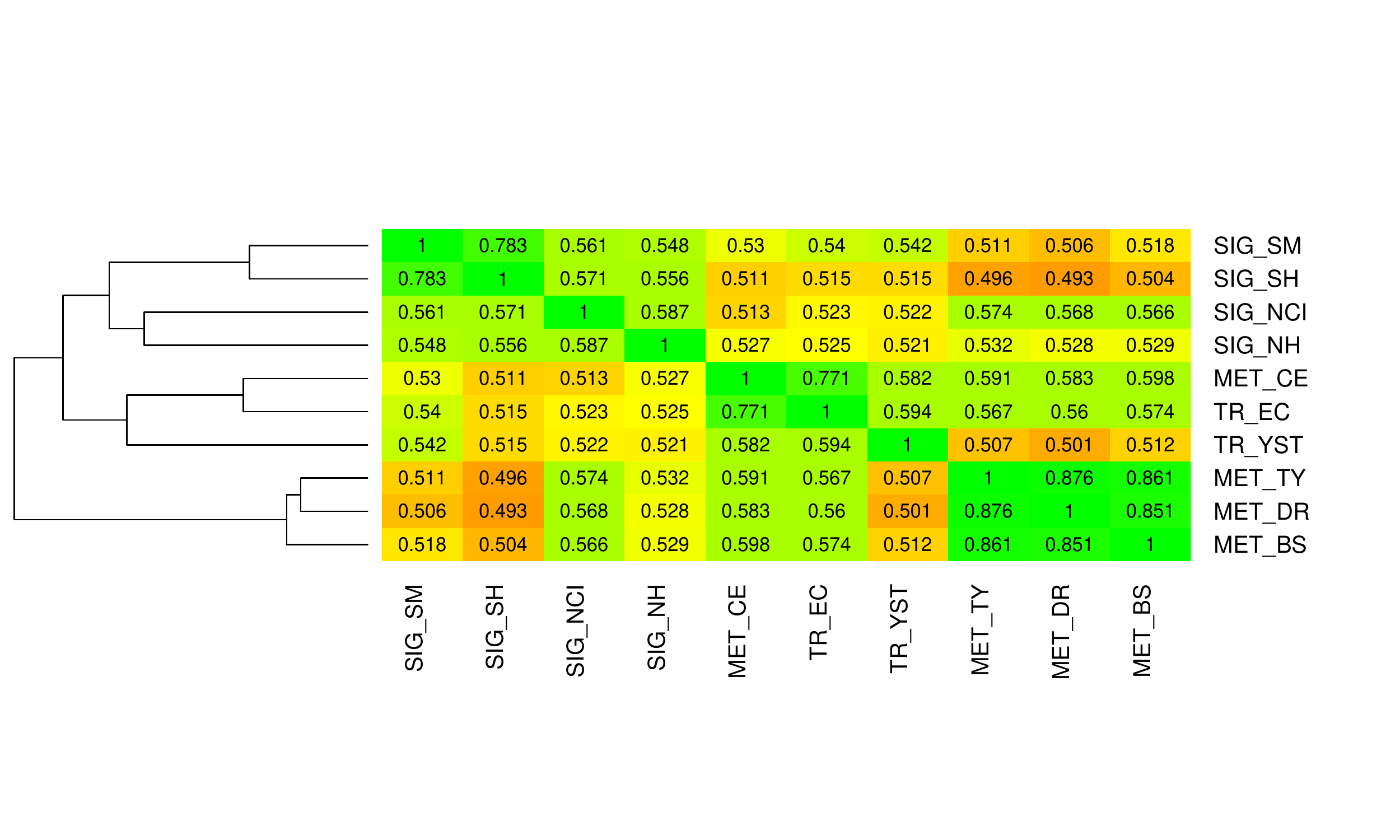}
   \vspace{-1cm}
   \caption{Heatmap and dendogram of the $dGDA_4$ between the tested networks. Cell signaling networks (\texttt{SIG}) and metabolic networks (\texttt{MET}) are clearly grouped together using directed graphlets information.}
   \label{fig:heatdir4}
\end{figure*}

A g-trie receives as input the list of graphs that the user wants to enumerate, which can be all undirected graphlets with up to five nodes, a set of directed graphs, specific interesting patterns (such as cliques or stars), or any other desired graphs. However, due to the nature of the g-trie which relies on common subtopologies between graphs to construct an efficient search tree, g-tries are better suited for tasks where one wants to count the occurrences of \textit{many small graphs} inside a large network.  G-Tries can be fully constructed before the enumeration and stored in a file or built on-the-fly to avoid having to store all possible graphs. 

\emph{Graphlet-tries} are an extension of g-tries that also consider the nodes' orbit. The broader term \emph{g-trie} is used whenever a concept applies to both g-tries and graphlet-tries. A graphlet-trie containing all the original 30 graphlets is shown in Figure~\ref{fig:graphlettrie}. 
The original orbit numbers are kept only for convenience since orbits are generated automatically in our implementation. An additional g-trie containing all non-bidirectional $d\mathcal{G}_4$ is shown in Figure~\ref{fig:graphlettrie_dir} (the bidirectional graphlets were removed for space concerns). 

Additionally, g-tries have already been extended to also support weighted \cite{choobdar2012motif} and colored \cite{ribeiro2014discovering} networks. The reader is refereed to \cite{Ribeiro2014} for more specific details on how a g-trie is created and used for subgraph enumeration. 

\section{Results and Discussion}

The experiments presented in this section can be divided into two categories: i) directed biological network comparison and ii) performance evaluation. The first illustrates how directed graphlets can be used to group a set of directed biological networks into different types according to their biological function, while the latter analyses the performance of our tool by comparing its execution time with state-of-the-art approaches on several biological networks.

Information relative to the networks used for testing purposes is summarized in Table~\ref{tab:networks}; the network set contains networks of different organisms and of three distinct types: cell signaling (prefix \texttt{SIG}), metabolic (\texttt{MET}) and transcriptional regulatory networks (\texttt{TR}). Their sizes vary from a few hundred nodes/edges to tens of thousand of nodes/edges.


\subsection{Directed Biological Network Comparison}

Different \emph{types} of biological networks such as metabolic, cell signaling and transcriptional regulatory networks pertain to distinct biological functions. The computational networks used in these experiments are evidently a translation of real biological networks, thus potentially making the process of finding their similarities harder since their real structure may not be fully represented. Nevertheless, it is usually assumed that network structural similarity of computational networks may also indicate functional similarity \cite{honey2010can}. In that case, networks belonging to the same type are expected to be more similar topologically than networks of different types.

In order to verify if that is the case for these specific networks we compare their directed graphlets so as to assess their topological similarity. This is done for each pair of networks $(G, H)$ from Table~\ref{tab:networks} by first enumerating all graphlet orbits and then comparing their $GDD$s by computing the $dGDA_k(G, H)$. These tests were performed for $u\mathcal{G}k$ and $d\mathcal{G}k$ with $k \in \left\{3, 4, 5\right\}$ but for space concerns only results for $d\mathcal{G}_4$ are presented. Furthermore, results from all experiments were found to be consistent with the results using $d\mathcal{G}_4$ for these particular networks but in general, different sets of (directed) graphlets can provide divergent information. 
	
The matrix obtained after computing the $dGDA_4(G,H)$ for each pair of networks is displayed in Figure~\ref{fig:heatdir4} along with the corresponding dendogram and heatmap. It can be observed that networks of the same type are correctly grouped by their $GDA$. This is an indicator that directed graphlets can detect topological similarities between directed biological networks of the same type and can likewise find structural differences between networks of different types.

\begin{table*}
\small
   \begin{minipage}{.28\linewidth}
      \begin{tabular}{|c|c|c|c|}
            \hline
            \multirow{2}{*}{$\mathcal{G}$} & \multirow{2}{*}{$|\mathcal{G}|$} & \multirow{2}{*}{$|\mathcal{O}|$} \\
            & &  \\
            \hline
            {$u\mathcal{G}_5$} & 30  &  73  \\
            \hline
            {$u\mathcal{G}_6$} & 142 &  480  \\
            \hline
            {$d\mathcal{G}_4$} & 214  &  730  \\
            \hline
            {$d\mathcal{G}_5$} & 9,578  &   45,637  \\
            \hline
            \multicolumn{3}{c}{\textit{(a)}}
            \end{tabular}
    \end{minipage}%
    \begin{minipage}{.18\linewidth}
    \centering
        \label{tab:simpleorbit}
            \begin{tabular}{|c|}
            \hline
            \multirow{2}{*}{\texttt{GraphCrunch}}  \\
             
               \\ \hline
            $6.59 \pm 2.60$  \\ \hline
            n/a \\ \hline
            n/a \\ \hline
            n/a \\ \hline
            \multicolumn{1}{c}{\textit{(b)}}
            \end{tabular}
    \end{minipage}
     \begin{minipage}{.21\linewidth}
        \centering
            \label{tab:eqorbit}
                \begin{tabular}{|c|}
                \hline
                \multirow{2}{*}{\texttt{ORCA}}  \\
                
                  \\ \hline
                 $2.08 \pm 1.15$ \\  \hline
                  n/a \\ \hline
                  n/a \\  \hline
                 n/a \\  \hline
                 \multicolumn{1}{c}{\textit{(c)}}
                \end{tabular}
        \end{minipage}
    \begin{minipage}{.24\linewidth}
    \centering
        \label{tab:noorbit}
            \begin{tabular}{|c|c|}
            \hline
            \multirow{2}{*}{\texttt{Kavosh}} & \multirow{2}{*}{\texttt{ESU}}\\
                &  \\
                        \hline
               $87.66 \pm 35.11$ & $73.00 \pm 30.81$ \\
            \hline
               $87.28 \pm 40.43$ & $70.91 \pm 31.98$ \\
            \hline
            
               $19.17 \pm 6.42$ & $17.33 \pm 5.99$ \\
            \hline
              $31.33 \pm 13.37$ & $28.60 \pm 11.79$ \\
            \hline
            
            \multicolumn{2}{c}{\textit{(d)}}
            \end{tabular}
    \end{minipage}
     \vspace{0.1cm}
\caption{Performance comparison between \texttt{gtrieScanner} and other algorithms. \textit{(a)} shows a description of the set of subgraphs being enumerated, as well as the total number of graphlets ($|\mathcal{G}|$) and orbits ($|\mathcal{O}|$). For a fair comparison we developed different versions of \texttt{gtrieScanner} that perform similar but distinct tasks. The speedups between our methods and other algorithms are shown in (b), (c) and (d). In \textit{(b)} the graphlet frequencies are obtained by enumerating all size $k$ graphlets. On the other hand, the variation in \textit{(c)} only enumerates the graphlets of up to size $k-1$ and then solves a set of equations to compute the frequencies of size $k$ graphlets. \texttt{GraphCrunch} and \texttt{ORCA} are only able to calculate undirected graphlets of up to size 5, so we can not compare their execution times with our own for $u\mathcal{G}_6$, $d\mathcal{G}_4$ or $d\mathcal{G}_5$. Finally, in \textit{(d)} we compare our approach to algorithms for network motif discovery, discarding orbit frequency computation. \label{tab:speedup}} 

\end{table*}

\begin{figure}[!ht]
   \centering
   \includegraphics[width=0.45\textwidth]{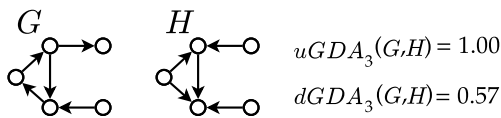}
   \caption{Two networks seen as identical using undirected graphlets that are 
   different if edge direction is considered. When enumerating $u\mathcal{G}_3$
   the $uGDA_3$ is equal to 1.00. On the other hand, if one uses $d\mathcal{G}_3$, the
   $dGDA_3$ is 0.57, substantially reducing their $GDA$. Directed graphlets are more appropriate to assess the level
   of similarity between directed networks since they capture more information about
   the networks' topologies.}
   \label{fig:gda}
\end{figure}

No explicit comparison with undirected graphlets is presented here since we do not claim that they are not capable of correctly grouping these specific networks into their respective classes. However, it is trivial to find cases in which undirected graphlets will not be able to distinguish networks that are clearly distinct when taking edge direction is taken into account (see Figure~\ref{fig:gda}).

\subsection{Performance Evaluation}

\subsubsection{Experimental Setting}

Our experimental results were gathered on a 8-core
machine consisting of two 4-core Intel® Xeon® Processor E5620
processors at 2.4GHz with a total of 12GB of memory. 
Code for \texttt{gtrieScanner} was developed in C++11 and compiled
using gcc 4.8.2. The other tools used for comparison were also
developed in C++ and are available as open-source. The execution times of each tool
are relative only to the graphlet enumeration phase, not taking into account
the time taken to load the graph into memory and perform other 
initialization and finalization tasks. 

\subsubsection{Performance Comparison}

The most used and well-known tool for graphlet discovery is \texttt{GraphCrunch} \cite{milenkovic2008graphcrunch}. Until recently it performed a full enumeration of all graphlets of up to size 5 in order to calculate their orbit frequency. Another tool, \texttt{ORCA} \cite{hovcevar2014combinatorial}, was shown to perform one or two orders of magnitude faster that the original \texttt{GraphCrunch} and has recently been integrated into it. Henceforth, when \texttt{GraphCrunch} is referenced we are alluding to its version before adopting \texttt{ORCA} to perform the enumeration. \texttt{ORCA} achieves its performance by observing that, given a limited set of graphs of size $k$, it is possible to build a system of equations to calculate their frequencies by using the frequencies of the size $k-1$ graphs and the frequency of a single graph of size $k$. This greatly reduces execution time since the size $k$ graphlet enumeration is much more computationally expensive than the size $k-1$ enumeration followed by solving the system of equations. While this approach brings substantial gains in terms of speed, it is quite limited in scope because a new set of equations needs to be manually derived in order to find the $k+1$ graphlets, $k+2$, and so on. These two tools also do not support edge direction, being only applicable for undirected graphlets.

Network motifs are similar in concept to graphlets and numerous tools for network motif discovery exist. Since tools for network motif discovery do not have to the calculate orbit frequencies specific to each node they perform less computational work than graphlet tools. Two well-known methods are \texttt{Kavosh} \cite{kashani2009kavosh} and \texttt{FANMOD} \cite{wernicke2006fanmod}, the latter being an implementation of the \texttt{ESU} algorithm; in our work a more efficient implementation of \texttt{ESU} \cite{ribeiro-PHD2011} is used. 


Because i) \texttt{GraphCrunch}, ii) \texttt{ORCA}, and iii) \texttt{ESU} and \texttt{Kavosh} methodologies are not comparable, three distinct versions of \texttt{gtrieScanner} were used: 
i) a version that enumerates all graphlets and orbits, ii) a version that enumerates size $k-1$ graphlets and orbits and then computes a set of equations to calculate the frequencies of the size $k$ graphlets and iii) a version that only enumerates the graphlets/network motifs and not the orbits, to be used respectively. The version that only enumerates motifs/graphlets was developed previous to this work and is available at {\small \verb|http://www.dcc.fc.up.pt/gtries/|}.

Results showing the average (mean) speedup obtained in all 10 networks for different sets of graphlets ($\mathcal{G}$) are presented in Table~\ref{tab:speedup}. Neither \texttt{GraphCrunch} nor \texttt{ORCA} are able to enumerate undirected graphlets with more than 5 nodes or directed graphlets. Our tool performs faster when compared to those two systems and, in addition to that, it can enumerate undirected graphlets of more than 5 nodes as well as directed graphlets. When compared to other tools for network motif discovery our tool is almost two orders of magnitude faster on average for undirected graphs and about 20 times faster for directed graphs. The speedups are lower for directed graphs because the search space is harder to constrain due to a higher number of graphlets and less common subtopologies between them. 

It is noticeable that the speedups of \texttt{gtrieScanner} are much higher when compared to network motif tools than to graphlet tools. This happens because tools for network motifs are general, meaning that they can be used for any subgraph/motif size, while tools to find graphlets are built to enumerate undirected graphlets of up to 5 nodes only. This allows them to have specialized optimizations that motif tools can not match.

From these experiments we can conclude that \texttt{gtrieScanner} can both perform faster than state-of-the-art graphlet tools and also provide a more general approach which supports any directed or undirected motif/graphlet size, as long as the full set of graphlets fits into memory.   


\section{Conclusions}


Recent advances in high-throughput cell biology caused huge amounts of cellular biological data to be continuously produced. Studying the large computational networks resulting from this information can lead to new insight into cellular organization. Due to the size of these networks it is necessary to resort to studying their smaller components such as Graphlets or like network motifs.

Graphlets in particular have been extensively applied to protein-protein interactions and other kinds of undirected networks but their applicability in directed biological networks, such as cell signaling, transcriptional regulatory and metabolic networks, is limited since they do not consider edge direction.  In this paper we highlight the importance of adapting graphlets to take into account edge direction by showing that networks of different types can be accurately grouped using directed graphlets. 

We present an efficient tool, \texttt{gtrieScanner}, that is able to compute directed and undirected graphlets of arbitrary size, as well as network motifs, as long as they fit into memory. Our tool also allows for the user to customize the set of graphs that he/she wants to query in the network, further demonstrating the flexibility of our tool. We assess our tool's performance on a set of directed biological networks and compare it to other tools for graphlet and network motif discovery. We observe that it is the fastest available tool for either task while also being a very general approach. Therefore, we believe that we have broadened the applicability of graphlets by extending them to directed graphlets and by providing a efficient tool  for that task.


\subsection*{Acknowledgements}

This work is partially funded by FCT (Portuguese Foundation for
Science and Technology) within project UID/EEA/50014/2013. David
Apar\'icio is supported by a FCT/MAP-i PhD research grant (PD/BD/105801/2014).

\bibliographystyle{IEEEtran}
\bibliography{refs.bib}

\end{document}